\newcommand{\mrm}{\mathrm}

\documentclass[vci-paper]{vciart}
\usepackage{graphicx}
\usepackage{amssymb}

\begin{document}

\begin{frontmatter}

\title{The ATLAS Level-1 Trigger: Status of the System and First
  Results from Cosmic-Ray Data}



\author[ROMA2]{G.~Aielli},
\author[KIP]{V.~Andrei},
\author[KIP]{R.~Achenbach},
\author[UL]{P.~Adragna},
\author[NAPOLI]{A.~Aloisio},
\author[NAPOLI]{M.G.~Alviggi},
\author[BOLOGNA]{S.~Antonelli},
\author[CERN]{S.~Ask},
\author[RUTHER]{B.M.~Barnett},
\author[MAINZ]{B.~Bauss}
% ...

\address[ROMA2]{Universita degli Studi di Roma "Tor Vergata" and INFN Roma II}
\address[KIP]{Kirchhoff-Institut für Physik, University of Heidelberg, D-69120 Heidelberg, Germany}
\address[UL]{Physics Department, Queen Mary, University of London, London E1 4NS, UK}
\address[NAPOLI]{Universita degli Studi di Napoli "Federico II" and INFN Napoli}
\address[BOLOGNA]{INFN Bologna and Università degli Studi di Bologna}
\address[CERN]{CERN, PH Department, Switzerland}
\address[RUTHER]{CCLRC Rutherford Appleton Laboratory, Chilton, Didcot, Oxon OX11 0QX, UK}
\address[MAINZ]{Institut für Physik, University of Mainz, D-55099 Mainz, Germany}
% ...

\begin{abstract}
The ATLAS detector at CERN's Large Hadron Collider (LHC) will be
exposed to proton-proton collisions from beams crossing at 40~MHz. At
the design luminosity of $10^{34}~\mathrm{cm}^{-2}~\mathrm{s}^{-1}$
there are on average 23 collisions per bunch crossing. A three-level
trigger system will select potentially interesting events in order to
reduce the read-out rate to about 200 Hz. The first trigger level is
implemented in custom-built electronics and makes an initial fast
selection based on detector data of coarse granularity. It has to
reduce the rate by a factor of $10^4$ to less than 100~kHz. The other
two consecutive trigger levels are in software and run on PC farms. We
present an overview of the first-level trigger system and report on
the current installation status. Moreover, we show analysis results of
cosmic-ray data recorded in situ at the ATLAS experimental site with
final or close-to-final hardware.
\end{abstract}
\end{frontmatter}

\section{Introduction}
A three-level trigger system was designed for the ATLAS experiment to
reduce the initial rate of 40~MHz to 100-200~Hz, a rate digestible by
the online-processing and archival-storage facilities. The trigger
system aims at retaining potentially interesting events by selecting
signatures of high-transverse-momentum particles or large missing
transverse energy. The first trigger level (LVL1)~\cite{L1TDR} of
ATLAS is based on data from the calorimeters and dedicated fast muon
detectors. It consists of the calorimeter trigger, the muon trigger,
and the central trigger system. The output rate of LVL1 is required to
be less than 100~kHz with an allowed latency of $2.5~\mu\mrm{s}$. The
remaining two trigger levels, the level-2 trigger (LVL2) and the event
filter, are implemented in software and run on commercial PC
farms~\cite{HLTTDR}. The LVL2 trigger design is a unique feature of
ATLAS. It is based on the concept of ``regions of
interests''~(RoIs). Algorithms request full-resolution data only from
a fraction of the detector, from regions that were identified by the
LVL1 trigger as \emph{regions of interest}. For that purpose, event
information like the coordinates of a particle candidate or energy and
momentum values are generated by the LVL1 trigger systems and sent to
the LVL2 trigger processors. Thereby the amount of full-resolution
data that is accessed by the LVL2 is less than $10\%$ of the total
event size significantly reducing the processing time. At LVL2 the
rate is reduced by about a factor of 50 to a few kHz. The event filter
in turn has access to the full resolution data of the whole
detector. It runs offline-like reconstruction and selection algorithms
and has to provide another factor of 10 in rate reduction to arrive at
the final storage rate of 100-200~Hz.

\section{The ATLAS LVL1 Trigger System}
\begin{figure}
 \centering
 \includegraphics[width=8cm,draft=false]{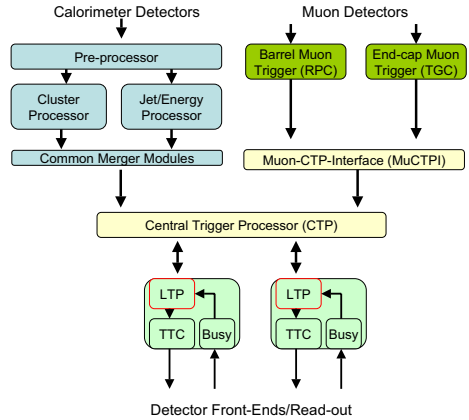}
 \caption{Sketch of the ATLAS LVL1 trigger system. Muon and
 calorimeter trigger detectors provide information about event
 candidates to the Central-Trigger Processor (CTP), which makes the
 LVL1 decision. If the event is accepted, a LVL1 accept signal is
 fanned out to all sub-detectors to initiate the readout. The LVL1
 trigger system generates (in addition to the readout data) regions of
 interest (RoIs) which are sent to the LVL2 trigger.}  
 \label{fig1}
\end{figure}

The first-level trigger system of ATLAS~(cf.\ Fig.~\ref{fig1})
synchronously processes information from the calorimeter and muon
trigger detectors at the heartbeat of the LHC, the proton-proton
bunch-crossing frequency of 40.08~MHz. It comprises three sub-systems,
the calorimeter trigger, the muon trigger, and the central-trigger
system.

\subsection{The LVL1 Calorimeter Trigger}
The LVL1 calorimeter trigger system receives trigger signals from the
calorimeter detectors, that is, the electromagnetic liquid-argon
calorimeter and the hadronic scintillator-tile
calorimeter. On-detector electronics combine the analogue signals to
7200 trigger towers, which are passed on to the calorimeter-trigger
system in the counting rooms next to the ATLAS cavern. The trigger
system consists of three subsystems. The
\emph{PreProcessor} receives the trigger-tower signals with a typical
granularity of $0.1 \times 0.1$ in $\eta$ and $\phi$. It digitises the
analogue signals, assigns a proton-proton bunch crossing to the
trigger pulses that extend over multiple bunch crossings, and does a
final lookup-table based calibration in transverse energy before
sending digital data to the next two calorimeter subsystems: the
algorithmic trigger processors. The \emph{Cluster Processor}
identifies and counts isolated electron/photon and hadron/tau lepton
candidates. The transverse energy of the electron/photon (hadron/tau)
candidates is discriminated against up to 16 (8) programmable
thresholds. The
\emph{Jet/Energy-sum Processor} identifies jet candidates and
discriminates them against 8 programmable thresholds. It also
calculates missing and total transverse energy of the whole event. 
\emph{Common-Merger Modules} carry out
merging before sending the data for each of the trigger objects
multiplicity and threshold information to the central-trigger system
synchronously with the 40~MHz machine. Upon reception of a LVL1-accept
signal (L1A) the data are sent to the readout system and the LVL2
trigger.

\begin{figure}
 \centering
 \includegraphics[width=8cm,draft=false]{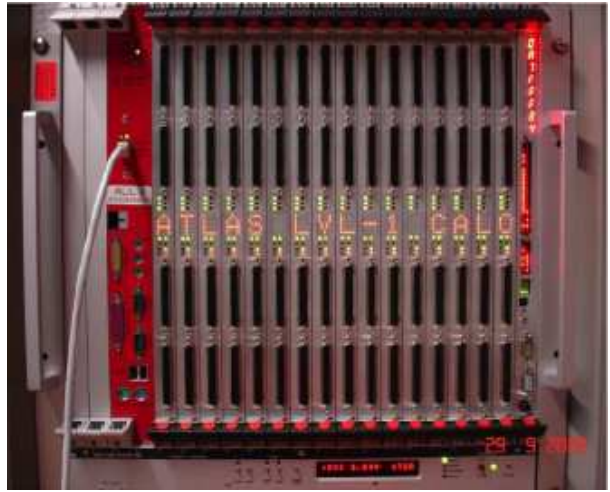}
 \caption{A full crate of pre-production PreProcessor modules of the
 LVL1 calorimeter trigger. Most of the functionality of these modules
 is on replaceable Multi-Chip Modules (MCMs). These MCMs each contain
 a processing ASIC and four ADCs. All the MCMs (3000) are available,
 final production of the 124 PreProcessor modules is in full swing and
 expected to be finished by March 2007.}
\label{fig2}
\end{figure}

\begin{figure}
 \centering
 \includegraphics[width=8cm,draft=false]{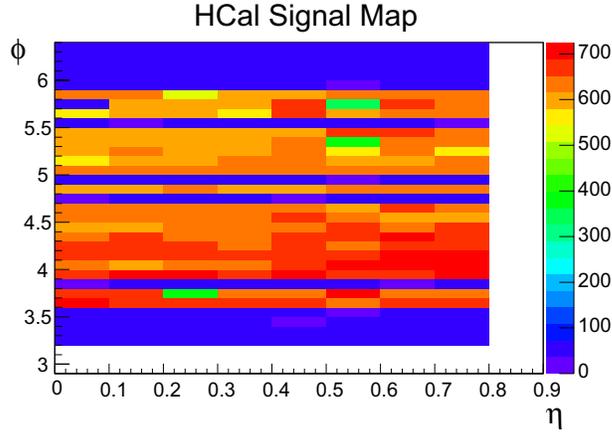}
 \caption{Map of calibration signals of one part of the hadronic
 calorimeter as measured by PreProcessor modules of the LVL1
 calorimeter trigger. The red and yellow colours are the expected
 signal range, green channels have a lower gain, and blue are pedestal
 values. The horizontal lines of channels with pedestals only stem
 from disconnected calorimeter electronics.}
 \label{fig3}
\end{figure}

The installation of the calorimeter-trigger system at the ATLAS
experimental site is proceeding well. Pre-production boards of the
three main components, the PreProcessor~(cf. Fig.~\ref{fig2}), the
Jet/Energy-sum Processor, and the Cluster Processor, are available and
tested. The production of the final boards is well advanced and
expected to be finished by the end of March 2007. The analogue trigger
cables connecting the calorimeter front-end electronics with the
off-detector calorimeter-trigger electronics are all produced and
tested, about 80\% are also installed and connected. First connection
tests with the central-trigger system were successful. Integration
with the calorimeters has started and is proceeding in parallel to the
commissioning of the trigger boards. Figure~\ref{fig3} shows a result
from these integration tests with the calorimeters. Using $1/8$ of the
barrel hadronic calorimeter, the connection was tested recording
calibration signals at the PreProcessor level. As can be seen from the
$\eta-\phi$ map shown in the figure, a reasonable signal was recorded
from almost all of the connected channels. One suspicious channel, at
$\eta=0 / \phi=5.8$, which was connected but saw no signal, is under
investigation. The connectivity tests proved also to be useful to
learn about signal shapes and noise levels.

\subsection{The LVL1 Muon Trigger}
The ATLAS muon spectrometer consists of muon chambers for precision
measurements and dedicated fast muon detectors for providing
information about muon candidates to the LVL1 Central Trigger. The
trigger chambers are resistive-plate chambers (RPCs) in the barrel
region $(|\eta|~<~1.05)$ and thin-gap chambers (TGCs) for the end-caps
$(1.05~<~|\eta|~<~2.4)$. The trigger detectors in both the barrel and
the end-caps are sub-divided in $\eta$ and $\phi$ space into trigger
sectors. In total there are 64 sectors for the barrel and 144 sectors
for the end-caps, the typical RoI size is $\Delta\eta \times
\Delta\phi = 0.1 \times 0.1$. 

The on-detector trigger electronics receives as input the pattern of
hits in the trigger chambers from more than $8 \times 10^5$
channels. Coincidences in different trigger stations are identified
independently in $\eta$ and $\phi$, based on geometrical roads whose
width is related to a programmable transverse-momentum threshold
making use of the deflection of muons in the magnetic field. The
coincidence allows six transverse momentum thresholds to be used at
the same time. Moreover, a proton-proton bunch crossing is assigned to
muon-candidate tracks. In the off-detector part of the muon trigger,
the \emph{Sector Logic}, which is situated in the counting rooms next
to the ATLAS cavern, combines the trigger data into muon candidates
per trigger sector. For each clock cycle, the muon trigger system
sends the muon candidate multiplicities per sector for each of the six
transverse-momentum thresholds to the central trigger.

\begin{figure}
 \centering
 \includegraphics[width=8cm,draft=false]{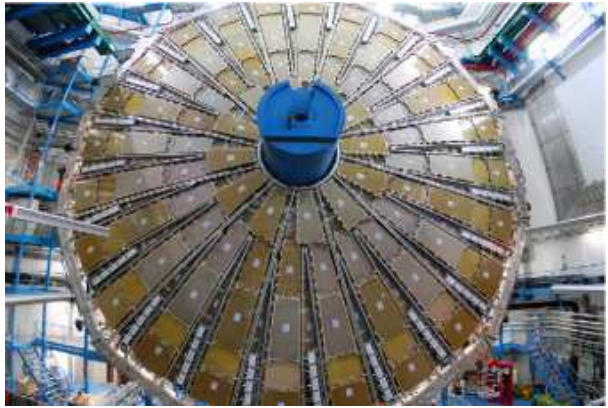}
 \caption{The first of the big wheels of the LVL1 muon-trigger end-caps. One
 can see the TGCs with on-detector electronics in
 the ATLAS cavern (October 2006).}
 \label{fig4}
\end{figure}

The installation of the LVL1 muon trigger chambers is in full
swing. The production of the final on-detector electronics is almost
finished:
\begin{itemize}
\item For the RPCs, all of the on-detector components are
installed. The barrel part of the muon spectrometer comprising the
chambers, the readout electronics and the services, entered the
commissioning phase and will be fully operational by November
2007. Pre-production versions of the off-detector parts of the LVL1
muon barrel trigger are already available and being used in
integration tests. Production of the final boards is expected to
commence in spring 2007.
\item For the TGCs, the chambers and on-detector electronics are
available and await installation in the ATLAS cavern. The end-cap part
of the muon spectrometer is organised in eight \emph{big wheels}, each
of which is 25~m in diameter. Six of the wheels are composed of
trigger chambers, while the other two are made of the precision
readout chambers (which are Monitored Drift Tubes). Currently
(February 2007), the first big TGC wheel is installed in the cavern,
cf.\ Fig.~\ref{fig4}, the second one is being
assembled. Pre-production versions of the Sector-Logic boards are
available and have been tested, mass production is about to commence
(March 2007). Further integration activities, especially with the
central-trigger system, have not yet begun but are planned for Spring
2007.
\end{itemize}

\subsection{The LVL1 Central Trigger}
The LVL1 Central Trigger is composed of the central-trigger processor
(CTP) and the Muon-to-CTP-Interface (MuCTPI). The MuCTPI receives muon
candidate information from the barrel and end-cap muon trigger
chambers and resolves cases where a candidate traverses a region with
overlapping trigger sectors to avoid double counting. It forms muon
multiplicities for the six configurable transverse-momentum thresholds
and sends these data to the CTP as trigger input. The CTP then derives
the LVL1 trigger decision based on the information received from the
calorimeter and muon trigger systems according to a programmable
trigger menu which aims at selecting high-transverse-momentum leptons,
photons and jets, as well as large missing and total transverse
energy. The CTP also applies prescale factors and dead time and
distributes the LVL1 accept (L1A) signal to the various sub-detectors
to initiate data readout. Busy signals from sub-detectors are
propagated to the CTP allowing to throttle the generation of L1As. For
accepted events the CTP and MuCTPI send data to the readout system as
well as to the LVL2 trigger.

A demonstrator of the MuCTPI and the final CTP system are already
installed in the underground counting room next to the ATLAS
cavern. The current MuCTPI provides almost the full functionality of
the final system, missing only some flexibility in the handling of
overlaps between muon trigger sectors. One sector input board out of
16 is currently available, corresponding to one octant of one half of
the detector. It provides inputs for 14 out of 208 trigger
sectors. The final boards are expected to become available for
integration in the experiment mid 2007. The CTP crate is already
equipped with the final boards. Both the CTP and MuCTPI are frequently
used in integration and commissioning activities at the ATLAS site.

\section{ATLAS Commissioning and Integration Activities}

\begin{figure}
 \centering
 \includegraphics[width=8cm,draft=false]{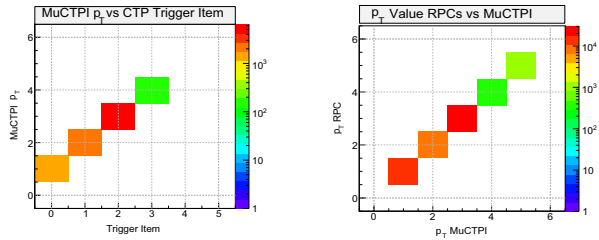}
 \caption{Plots from the combined cosmic-ray runs of the muon-barrel
 detector and the central-trigger system. The left-hand side shows the
 correlation between the muon-candidate transverse-momentum threshold
 ($p_T$) from the MuCTPI data and the trigger item fired in the
 CTP. On the right-hand side, the correlation between the muon $p_T$
 from the RPC data and the value from the MuCTPI data is plotted.}
 \label{fig5}
\end{figure}

Various combined cosmic-ray runs including different sub-detectors of
ATLAS were performed since August 2006. For some of these test runs,
temporary gas systems were operational for one sector of the muon
barrel detector~\cite{RPC}, comprising six precision readout chambers
(Monitored Drift Tubes) and six RPC trigger towers. The on-detector
coincidence boards were connected to one off-detector sector-logic
board which determined muon candidates per transverse-momentum
threshold and sent them to the central trigger. The central trigger in
turn was configured to trigger on any muon candidate and provided the
L1A that initiated the readout of the sub-detectors. The data was used
for validation in all involved sub-systems. Figure~\ref{fig5} shows
correlation plots of data that was transmitted from the LVL1
muon-trigger electronic to the MuCTPI, and from the MuCTPI to the
CTP. On the left-hand side, we plot the trigger item fired on the CTP
versus the transverse-momentum threshold at the MuCTPI, on the
right-hand side the candidate's momentum threshold from the RPC data
versus the MuCTPI data. As desired, there is an exact one-to-one
correlation in both plots demonstrating the data integrity.

\begin{figure}
 \centering
 \includegraphics[width=8cm,draft=false]{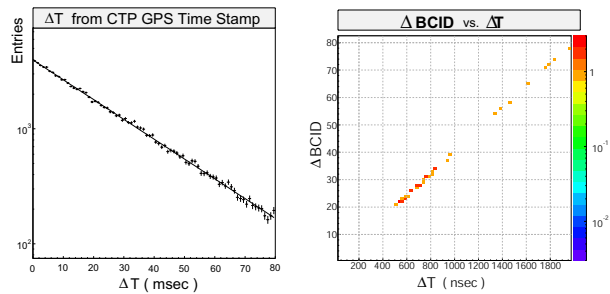}
 \caption{Plots determined from CTP data from cosmic-ray runs. On the
 left-hand side the time difference between consecutive events
 ($\Delta$~T) is shown. An exponential fit (black line) describes the
 data very well. To validate the GPS-time-stamp assignment, we plot
 the correlation of the difference of the bunch-crossing identifier
 and the time difference for consecutive events on the right-hand
 side.}  \label{fig6}
\end{figure}

The left-hand side of Fig.~\ref{fig6} shows the distribution of time
differences between consecutive events, determined from the CTP time
stamp. As expected, the events are exponentially distributed. The
right-hand side of Fig.~\ref{fig6} verifies the correct functioning of
the GPS-time-stamp assignment to each event. The plot shows the
correlation between the difference in the bunch-crossing identifier
(BCID) of consecutive events and the GPS-time difference of these
events. As expected from the fixed bunch-crossing frequency (of
40~MHz), there is a one-to-one correlation between the two
quantities. One can also see that there are no events closer together
than 20 bunch crossings, which is expected from the dead-time settings
in the CTP for this particular run (for every L1A the CTP introduced a
dead-time of 20 bunch crossings).

An ATLAS-wide effort to establish regular cosmic-ray data taking
activities to integrate step by step the whole of ATLAS towards first
collisions at the end of 2007 was started in December 2006. Cosmic
rays were recorded with parts of the barrel hadronic and
electromagnetic calorimeters. The CTP provided a common clock and
distributed the L1As to initiate the detector readout. The next
campaign beginning in March 2007 will include besides the barrel
calorimeters also parts of the barrel muon detectors.

Another important milestone for the trigger system of ATLAS was
reached in February 2007 with the successful integration of a full
muon trigger slice up to LVL2. A LVL2 trigger algorithm was run for
the first time in real online mode during a cosmic-ray run. One sector
of the LVL1 muon barrel trigger was operated together with MuCTPI and
CTP, which provided RoI data to the LVL2 farm. A trigger algorithm was
run that selected downward-going muon candidates. The algorithm was
seeded by the LVL1 RoI data and requested in turn sub-detector data
from the readout system to reconstruct a muon candidate track. The
algorithm was found to run as expected, reconstructing muon candidates
with a mean processing time of 1~millisecond, well within the LVL2 time
budget of 10~milliseconds.

\section{Conclusions}
The installation and integration of the ATLAS LVL1 trigger system is
now in full swing. The final modules are mostly already produced and
are being installed in the experiment as they become available. Parts
of the system, especially one sector of the muon-barrel trigger and
the central-trigger system, are frequently operated in test runs in
combination with other ATLAS sub-detectors like the barrel
calorimeters. These cosmic-ray data are used for system validation and
data integrity checks. On the LVL1 trigger side, integration of the
calorimeter-trigger system and the end-cap muon trigger system with
the central trigger are expected to commence in spring 2007.

\clearpage
\clearpage


\end{document}